\documentclass[twocolumn,showpacs,superscriptaddress,floatfix,prb]{revtex4}
\usepackage{graphicx,amsfonts,amssymb,amsmath,hyperref,hypcap,enumerate}

\begin{document}
\title{Moir\'e Assisted Fractional Quantum Hall State Spectroscopy}

\author{Fengcheng Wu}
\affiliation{Department of Physics, University of Texas at Austin, Austin, TX 78712, USA}
\affiliation{Materials Science Division, Argonne National Laboratory, Argonne, IL 60439, USA}

\author{A.H. MacDonald}
%\email{macdpc@physics.utexas.edu}
\affiliation{Department of Physics, University of Texas at Austin, Austin, TX 78712, USA}

\date{\today}

\begin{abstract}
Intra-Landau level excitations in the fractional quantum Hall regime are not accessible via optical
absorption measurements.  We point out
that optical probes are enabled by the periodic potentials produced by a moir\'e pattern.
Our observation is motivated by the recent observations of fractional quantum Hall 
incompressible states in moir\'e-patterned graphene on 
a hexagonal boron nitride substrate, and is theoretically 
based on $f-$sum rule considerations supplemented by  
a perturbative analysis of the influence of the moir\'e potential on many-body states.
 \end{abstract}

%73.43.-f Quantum Hall effects
%73.22.Pr Electronic structure of graphene
% 78.67.-n Optical properties of low-dimensional, mesoscopic, and nanoscale materials and structures

\maketitle

\section{Introduction}

When electrons in two-dimensions partially occupy a
macroscopically degenerate Landau level (LL), the character of the ground state and of its excitations
both change in a complex way as a function of filling factor $\nu$ and LL kinetic energy index $n$.  
The charged excitation gaps (chemical potential discontinuities) that appear at 
many rational LL filling factors are efficiently exposed by transport measurements 
because\cite{LesHouches} they give rise to fractional quantum Hall (FQH) effects.
It has however been a stumbling block in explorations of FQH physics that many other aspects of the uniquely subtle many-electron 
states are hidden from view, and in particular that excitations within a single LL of a 
two-dimensional electron system (2DES) are optically dark.  
In this article we propose an approach which can be used to 
make them visible. 

When two van der Waals materials form a heterojunction, misalignment and lattice constant differences
give rise to a periodic moir\'e pattern that makes all local observables periodic functions of position.
Moir\'e patterns are particularly important when 
formed in graphene sheets because the high quality of 
these 2DESs helps make their influence dominate over random inhomogeneity induced by uncontrolled disorder.  
Moir\'e patterns formed in graphene on hexagonal boron nitride (hBN) and graphene on 
graphene have recently \cite{Hunt13, Dean13, Yu14, Dean15, TutucNew} been successfully used to realize 
Hofstader butterfly systems with fractal band spectra that 
are extraordinarily sensitive to commensurability between magnetic-field and 
periodic potential area scales.  Here we show that they also enable coupling between
light and intra-LL collective excitations.

To illustrate our ideas we focus on the collective excitations of the strongest fractional
quantum Hall incompressible states, those that occur at $\nu=1/3$ and $\nu=2/3$ which were 
first understood by Laughlin and are named in his honor.
The collective excitations of these states are accurately 
described by the single-mode approximation of Girvin, MacDonald and Platzman\cite{GMP}. 
Because of analogies between these excitations and the 
roton modes in superfluid helium, collective excitation of FQH states are
known as magneto-rotons. 
Magneto-roton properties have been investigated using a variety of approaches, for example by    
using exact diagonalization\cite{Haldane, Jolicoeur16} methods or 
applying composite bosons\cite{Zhang91} or composite fermions\cite{Jain92} ideas, 
and continue to be actively studied.  Recent advances include the identification of a 
connection to Hall viscosity\cite{Haldane09,Haldane12} and an analysis of their relationship to 
the stability of FQH states\cite{Sondhi13}. 

The experimental observation of the intra-LL collective excitations of FQH states has 
been an ongoing challenge because of the absence in homogeneous fluids  
of dipole coupling between light and any intra-LL neutral excitation.  
Inelastic light scattering has provided indirect signatures of intra-LL collective 
excitations\cite{West93, West01, West05, West13} which are thought to be enabled by disorder
which breaks translational symmetry and enables coupling between
light and finite-momentum excitations, but does not allow for momentum resolution.
We show below that weak moir\'e patterns expose 
collective excitations only at the moir\'e pattern reciprocal lattice wave vectors, which can be 
tuned by varying the van der Waals heterojunction twist angle. 

Our paper is organized as follows. In Sec.~\ref{f-sum-rule}, we derive a strong magnetic field 
$f-$sum rule and use it to show quite generally that the contribution of intra-LL excitation to the optical conductivity is finite 
in the presence of a spatially varying potential.  In Sec.~\ref{perturbation}, we use perturbation theory 
to account for the influence of moir\'e potential on incompressible FQH states. 
The perturbative approach is valid when 
the periodic moir\'e potential is weak
compared to the collective mode excitation energies. 
In Sec.~\ref{SMA}, we also make a single mode approximation
to provide an explicit expression for the optical conductivity of the $\nu=1/3$ FQHE states. 
Finally in Sec.~\ref{discussion}, we discuss possible experimental
systems, including graphene/hBN  and twisted transition metal 
dichalcogenides (TMD) bilayers.

\section{Strong Magnetic Field $f-$sum rule}
\label{f-sum-rule}
\begin{figure}[t]
	\includegraphics[width=0.9\columnwidth]{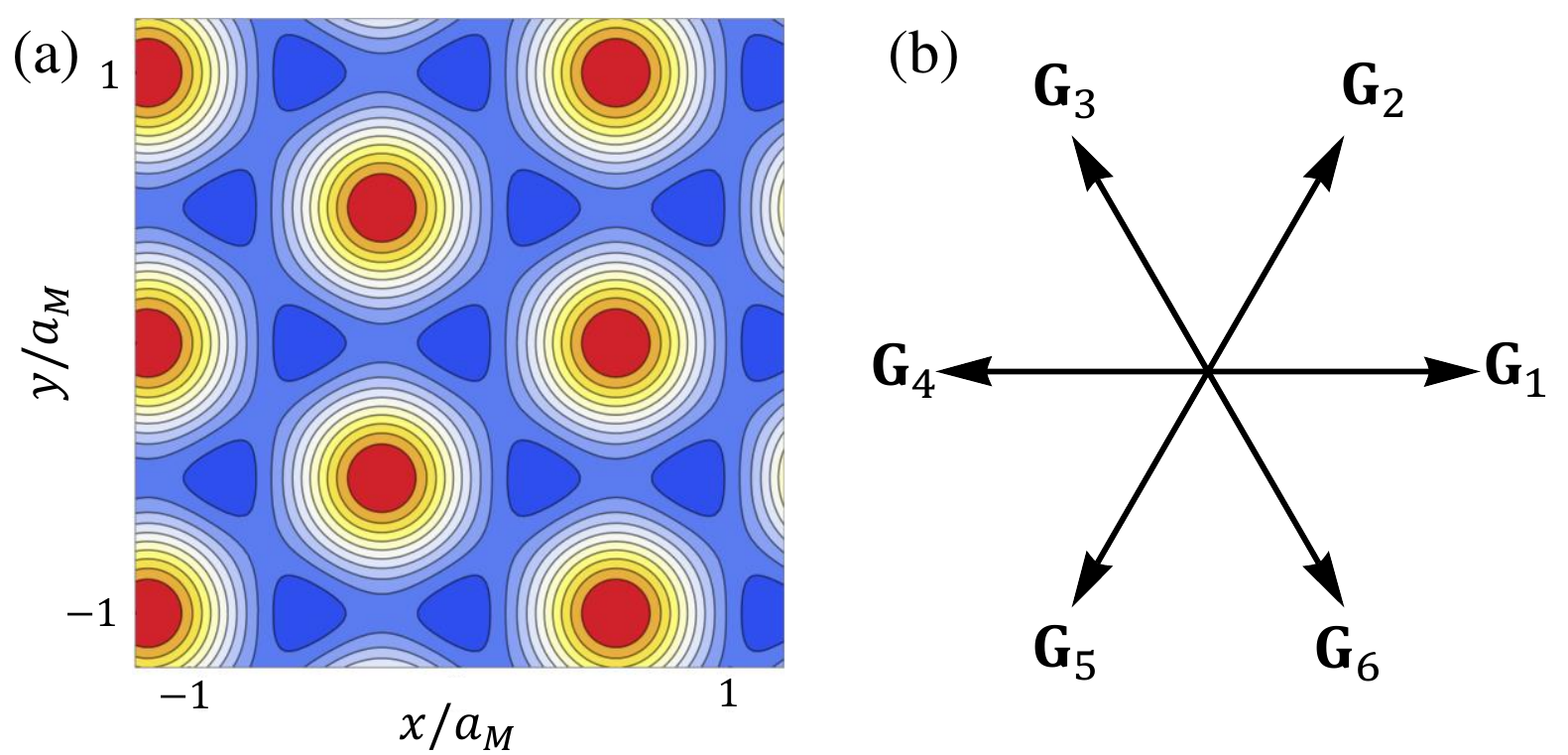}
	\caption{(Color online)(a) Schematic illustration of a periodic potential in real space due to
	a moir\'e pattern formed between 2D crystals with triangular Bravais lattices. 
	(b) The first shell of moir\'e reciprocal lattice vectors.}
	\label{Fig:Illustration}
\end{figure}

We consider electrons in the lowest-LL and for the moment neglect possible spin or valley degrees of freedom. 
When projected to the lowest LL, 
the Hamiltonian includes only Coulomb interaction $\hat{H}_C$ and moir\'e potential $\hat{V}$ terms, 
and is given up to a constant by:
\begin{equation}
\begin{aligned}
\hat{H} &= \hat{H}_C+\hat{V},\\
\hat{H}_C &=\frac{1}{2}\int \frac{d^2\bold{q}}{(2\pi)^2} v_C(\bold{q}) \;  \bar{\rho}_{-\bold{q}}\,\bar{\rho}_{\bold{q}},\\
\hat{V} &=\sum_{\bold{G}} V_{\bold{G}}\bar{\rho}_{\bold{G}},
\end{aligned}
\label{Ham}
\end{equation}
where 
\begin{equation}
\bar{\rho}_{\bold{q}} =\sum_{j}\overline{\text{exp}[-i\bold{q}\cdot\bold{r}_j]}
\end{equation}
is the LL projected density operator,
and $v_C(\bold{q})=2\pi e^2/(\varepsilon |\bold{q}|)$ is the Coulombic electron-electron interaction.
The potential $\hat{V}$ is produced by the moir\'e pattern, and is periodic as
illustrated in Fig.~\ref{Fig:Illustration}(a). 
For graphene/hBN\cite{sublattice} the spatial variation of 
$\hat{V}$ is accurately characterized by a Fourier expansion that includes only 
the six wave vectors in the first shell of moir\'e reciprocal lattice shown in Fig.~\ref{Fig:Illustration}(b).\cite{Bistritzer11} 
The summation over $\bold{G}$ in Eq.~(\ref{Ham}) is restricted to these six vectors. 
Because the potential is real and the moir\'e pattern has \cite{Falko13} three-fold rotational symmetry, we have the following constraints:
\begin{equation}
V_{\bold{G}_1}=V_{\bold{G}_3}=V_{\bold{G}_5}=V_{\bold{G}_2}^*=V_{\bold{G}_4}^*=V_{\bold{G}_6}^*.
\end{equation}
The magnitude of $\bold{G}$ in the first shell can be varied by 
adjusting the twist angle $\theta$. 
For small $\theta$:
\begin{equation}
|\bold{G}|  = \frac{4\pi}{\sqrt{3}a_M} \; , \; a_M \approx a/\sqrt{x^2+\theta^2},
\end{equation}
where $a_M$ is the moir\'e periodicity,  $ x = |a'-a|/a$, and $a$ and $a'$ are the lattice constants of the two layers\cite{Jung14}.

The optical conductivity $\sigma(\omega)$ of a material can be probed 
by measuring optical reflection, transmission, or absorption.  
Theoretically, the longitudinal conductivity can be related to the density-density response 
function $\chi$ using 
\begin{equation}
\sigma(\bold{q},\omega)=e^2 \frac{i\omega}{|\bold{q}|^2} \Pi(\bold{q},\omega),
\label{s.p}
\end{equation}
where the polarization function $\Pi$ satisfies 
\begin{equation}
\Pi^{-1}(\bold{q},\omega) =  v_C(\bold{q})+ \chi^{-1}(\bold{q},\omega) 
\label{c.p}
\end{equation}
Eq.(\ref{s.p}) follows from the definition of the conductivity as the current response to 
internal electric field, and from the charge continuity equation. 
%Eq.(\ref{c.p}) can be taken as the definition of polarization function.
We introduce the dynamic structure factor $S$:
\begin{equation}
S(\bold{q},\epsilon)=\frac{1}{N}\sum_{m>0} |\langle \Psi_m |\bar{\rho}_{\bold{q}}| \Psi_0 \rangle|^2 \delta(\epsilon - (E_m -E_0)),
\label{dynamic_str}
\end{equation}
where $N$ is the number of electrons, and $|\Psi_m\rangle$ and $E_m$ are the exact 
many-body eigenfunctions and eigenvalues of the many-body Hamiltonian, 
and the label $m=0$ is reserved for the ground state.
The density response function $\chi$ can be expressed in terms of $S$:
\begin{equation}
\begin{aligned}
\text{Im}\chi(\bold{q},\omega)&=-\frac{\pi N}{A}  [S(\bold{q},\hbar \omega)-S(-\bold{q},-\hbar \omega)],\\
\text{Re} \chi(\bold{q},\omega)&=-\frac{1}{\pi}\int_{-\infty}^{+\infty}\frac{\text{Im}\chi(\bold{q},\omega')}{\omega-\omega'+i0^+}d\omega',
\end{aligned}
\label{Imchi}
\end{equation}
where $A$ is the area of the system, and the second equation follows from Kramers-Kronig relations.

It follows from Eqs.(\ref{dynamic_str}) and (\ref{Imchi}) that 
in a system with an energy gap both $S(\bold{q},\epsilon)$ and $\chi(\bold{q},\omega)$ 
vanish at least as fast as $|\bold{q}|^2$ at small $|\bold{q}|$.
From Eq.(\ref{c.p}), we conclude that to this order $\Pi(\bold{q},\omega) \approx \chi(\bold{q},\omega) $ and that  
\begin{equation}
\sigma(\omega) 
%=\lim_{|\bold{q}| \to 0}\sigma(\bold{q},\omega)=ie^2 \omega \lim_{|\bold{q}| \to 0}\frac{\Pi(\bold{q},\omega)}{|\bold{q}|^2} \\
=ie^2 \omega \lim_{|\bold{q}| \to 0}\frac{\chi(\bold{q},\omega)}{|\bold{q}|^2}.
\end{equation}
In particular for the real part of the optical conductivity which is responsible for optical absorption
\begin{equation}
\text{Re}\sigma(\omega)=\frac{\pi e^2 N}{A}  \lim_{|\bold{q}| \to 0} \frac{\omega S(\bold{q},\hbar\omega)}{|\bold{q}|^2},\; \omega>0.
\label{Resigma}
\end{equation}

It is instructive to evaluate the first moment $\bar{f}(\bold{q})$ of $S(\bold{q}, \epsilon)$. $\bar{f}$ 
does not require knowledge of any many-body eigenstates, but provides valuable insights into 
the behavior of the response functions $\sigma$ and $\chi$.
\begin{equation}
\bar{f}(\bold{q}) = \int_{0}^{+\infty} \epsilon S(\bold{q}, \epsilon) d\epsilon = \frac{1}{N} \langle \bar{\rho}_{-\bold{q}} [\hat{H}, \bar{\rho}_{\bold{q}}] \rangle_0, 
\end{equation}
where $\langle ... \rangle_0$ denotes an expectation value in the ground state $|\Psi_0 \rangle$. 
Since spatial inversion symmetry can be broken by the potential $\hat{V}$, we define the symmetrized first moment:
\begin{equation}
\bar{f}_+(\bold{q}) \equiv (\bar{f}(\bold{q})+\bar{f}(-\bold{q}))/2=\frac{1}{2N}\langle [\bar{\rho}_{-\bold{q}}, [\hat{H}, \bar{\rho}_{\bold{q}}]] \rangle_0.
\end{equation}

The contributions of $\hat{H}_C$ and potential $\hat{V}$ to $\bar{f}_+(\bold{q})$ can be evaluated separately:
\begin{equation}
\begin{aligned}
&\bar{f}_+(\bold{q}) = \bar{f}_+^{(C)}(\bold{q})+\bar{f}_+^{(V)}(\bold{q}),\\
&\bar{f}_+^{(C)}(\bold{q})=\frac{1}{2N}\langle [\bar{\rho}_{-\bold{q}}, [\hat{H}_C, \bar{\rho}_{\bold{q}}]] \rangle_0 ,\\
&\bar{f}_+^{(V)}(\bold{q})=\frac{1}{2N}\langle [\bar{\rho}_{-\bold{q}}, [\hat{V}, \bar{\rho}_{\bold{q}}]] \rangle_0 .
\end{aligned}
\end{equation}

%As demonstrated by Girvin-MacDonald-Platzman\cite{GMP}, $\bar{f}_+^{(C)}(\bold{q})$ vanishes at order $|\bold{q}|^2$. 
%We reproduce the argument here.  
$\bar{f}_+^{(C)}(\bold{q})$ can be expressed as follows\cite{LesHouches}:
\begin{equation}
\begin{aligned}
\bar{f}_+^{(C)}(\bold{q}) = \int \frac{d^2\bold{k}}{(2\pi)^2}& v_C(\bold{k})\big[1- \cos(\hat{z}\cdot(\bold{k}\times \bold{q}))\big]\\
\times&\big[\tilde{\bar{s}}(\bold{k}+\bold{q})-\tilde{\bar{s}}(\bold{k}) \big]e^{-|\bold{k}|^2/2},
\end{aligned}
\label{fCoulomb}
\end{equation}
where $\tilde{\bar{s}}(\bold{k}) = \exp(|\bold{k}|^2/2) \bar{s}(\bold{k})$. $\bar{s}(\bold{k})$ is the static structure factor with respect to the ground state of the full Hamiltonian $\hat{H}$:
\begin{equation}
\bar{s}(\bold{k})=\frac{1}{N} \langle \Psi_{0} | \bar{\rho}_{\bold{q}}^\dagger \bar{\rho}_{\bold{q}}  |  \Psi_{0} \rangle.
\end{equation}
In Eq.~(\ref{fCoulomb}), the factor $\big[1- \cos(\hat{z}\cdot(\bold{k}\times \bold{q}))\big]$ scales as $|\bold{q}|^2$, while the other factor $\big[\tilde{\bar{s}}(\bold{k}+\bold{q})-\tilde{\bar{s}}(\bold{k}) \big]$ vanishes at $\bold{q}=0$. Since $\bar{f}_+^{(C)}(\bold{q})$ is an even analytic function of $\bold{q}$ as long as excitation gap is finite, it follows that its leading long-wavelength behavior is $\sim \bold{q}^4$ even in the presence of the moir\'e perturbation.\cite{GMP}

On the other hand $\bar{f}_+^{(V)}(\bold{q})$ is finite at second order
 in $|\bold{q}|$, as shown below:
\begin{equation}
\begin{aligned}
&\bar{f}_+^{(V)}(\bold{q})=\frac{1}{2N}\sum_{\bold{G}}V_{\bold{G}}\langle [\bar{\rho}_{-\bold{q}}, [\bar{\rho}_{\bold{G}}, \bar{\rho}_{\bold{q}}]] \rangle_0  \\
=&\frac{1}{2N}\sum_{\bold{G}}V_{\bold{G}} \big[(e^{\ell_B^2G^*q/2}-e^{\ell_B^2q^*G/2})\\
&\times(e^{-\ell_B^2q^*G/2}-e^{-\ell_B^2G^*q/2})e^{-\ell_B^2q^*q/2}\big] \langle \bar{\rho}_{\bold{G}} \rangle_0\\
\approx & -\frac{1}{2N}\sum_{\bold{G}} [\ell_B^2(\bold{q} \times \bold{G}) \cdot \hat{z}]^2 V_{\bold{G}}\langle \bar{\rho}_{\bold{G}} \rangle_0,
\end{aligned}
\label{fsum}
\end{equation}
where  magnetic length $\ell_B$ is $\sqrt{\hbar/(eB)}$  and $q=q_x+iq_y$. $\langle \bar{\rho}_{\bold{G}} \rangle_0$ is finite due to the potential $\hat{V}$.  Eq.(\ref{fsum}) relies on the well-known\cite{GMP} commutation relations of LL projected
density operators.   Since $\bar{f}(\bold{q})$ equals $\bar{f}(-\bold{q})$ up to second order in $|\bold{q}|^2$ by definition (see Eq.(\ref{dynamic_str})),
%$\bar{f}(\bold{q}) \approx \bar{f}_+(\bold{q}) \approx \bar{f}_+^{(V)}(\bold{q}) $, which is again correct up to $O(|\bold{q}|^2)$. Therefore, 
we obtain the following $f-$sum rule:
\begin{equation}
\begin{aligned}
\int_{0}^{+\infty} \text{Re}\sigma(\omega) d(\hbar \omega) =  -\frac{e^2}{\hbar}\frac{1}{8 N_\phi}\sum_{\bold{G}}\ell_B^2 |\bold{G}|^2 V_{\bold{G}}\langle \bar{\rho}_{\bold{G}} \rangle_0,
\label{sum-rule}
\end{aligned}
\end{equation}
%\begin{widetext}
%\begin{equation}
%\begin{aligned}
%&\int_{0}^{+\infty} \epsilon S(\bold{q}, \epsilon) d\epsilon \\=& -\frac{1}{2N}\sum_{\bold{G}}[\ell_B^2(\bold{q} \times \bold{G}) \cdot \hat{z}]^2 V_{\bold{G}}\langle \bar{\rho}_{\bold{G}} \rangle_0+O(q^3),\\
% \int_{0}^{+\infty}\text{Re}\sigma(\omega) d(\hbar \omega)  =   -\frac{e^2}{\hbar}\frac{1}{4 N_\phi}\sum_{\bold{G}}[\ell_B(\hat{q} \times \bold{G}) \cdot \hat{z}]^2 V_{\bold{G}}\langle \bar{\rho}_{\bold{G}} \rangle_0,
%\end{aligned}
%\label{sum-rule}
%\end{equation}
%\end{widetext} 
where $N_\phi=A/(2\pi l_B^2)$ is the LL degeneracy.  The final form for Eq.(\ref{sum-rule}) assumes
that $V_{\bold{G}}$ has a three-fold rotational symmetry so that the longitudinal conductivity tensor is 
isotropic.  This sum rule proves that light is absorbed by intra-LL excitations
when a moir\'e pattern is established.

\section{Perturbation theory}
\label{perturbation}
To gain deeper insight we assume that the potential $|V_{\bold{G}}|$ is small compared to the Coulomb interaction energy scale $e^2/(\varepsilon \ell_B)$ and apply perturbation theory.
We denote the eigenstates and eigenenergies of the projected Coulomb interaction $H_C$ respectively by $|\Psi_{\bold{k},m}^{(0)}\rangle$ and $E_{\bold{k},m}^{(0)}$, where $\bold{k}$ is the total 
momentum quantum number of a many-body state\cite{Haldane85} and $m$ distinguishes states at the same $\bold{k}$. 
For filling factors at which the fractional quantum Hall effect occurs, the Coulomb ground state is translationally 
invariant and we denote it by $|\Psi_{0}^{(0)}\rangle$.
Treating the potential $\hat{V}$ as a weak perturbation, we obtain at first-order in $|V_{\bold{G}}|$:
\begin{equation}
\begin{aligned}
|\Psi_{\bold{k},m}\rangle &\approx |\Psi_{\bold{k},m}^{(0)}\rangle + |\Psi_{\bold{k},m}^{(1)}\rangle, \\
|\Psi_{\bold{k},m}^{(1)}\rangle & = \sum_{\bold{G}, n} \frac{V_{\bold{G}} \langle \Psi_{\bold{k}+\bold{G},n}^{(0)} |\bar{\rho}_{\bold{G}} | \Psi_{\bold{k},m}^{(0)} \rangle}{E_{\bold{k},m}^{(0)}-E_{\bold{k}+\bold{G},n}^{(0)}}
  | \Psi_{\bold{k}+\bold{G},n}^{(0)} \rangle.
\end{aligned}
\end{equation} 

We work out the matrix element for the projected density operator to first order in $V_\bold{G}$:
\begin{equation}
\begin{aligned}
&\langle \Psi_{\bold{k},m} |\bar{\rho}_{\bold{q}}| \Psi_{0} \rangle \\
\approx & \delta_{\bold{k},\bold{q}} \langle \Psi_{\bold{q},m}^{(0)} |\bar{\rho}_{\bold{q}}| \Psi_{0}^{(0)} \rangle
+ \langle \Psi_{\bold{k},m}^{(0)} |\bar{\rho}_{\bold{q}}| \Psi_{0}^{(1)} \rangle
+ \langle \Psi_{\bold{k},m}^{(1)} |\bar{\rho}_{\bold{q}}| \Psi_{0}^{(0)} \rangle\\
\approx & \delta_{\bold{k},\bold{q}} \langle \Psi_{\bold{q},m}^{(0)} |\bar{\rho}_{\bold{q}}| \Psi_{0}^{(0)} \rangle\\
& + \sum_{\bold{G}, n} \delta_{\bold{k},\bold{q}+\bold{G}} 
\frac{V_{\bold{G}} \langle \Psi_{\bold{G},n}^{(0)} |\bar{\rho}_{\bold{G}} | \Psi_{0}^{(0)} \rangle}{E_{0}^{(0)}-E_{\bold{G},n}^{(0)}}
\langle \Psi_{\bold{q}+\bold{G},m}^{(0)} |\bar{\rho}_{\bold{q}}| \Psi_{\bold{G},n}^{(0)} \rangle \\
& + \sum_{\bold{G}, n} \delta_{\bold{k},\bold{q}+\bold{G}} 
\frac{V_{\bold{G}} \langle \Psi_{\bold{q}+\bold{G},m}^{(0)} |\bar{\rho}_{\bold{G}} | \Psi_{\bold{q},n}^{(0)} \rangle}{E_{\bold{q}+\bold{G},m}^{(0)}-E_{\bold{q},n}^{(0)}}
\langle \Psi_{\bold{q},n}^{(0)} |\bar{\rho}_{\bold{q}}| \Psi_{0}^{(0)} \rangle.
\end{aligned}
\label{den_ME}
\end{equation}

To evaluate the conductivity, we need to retain only terms up to first order in $|\bold{q}|$ 
in Eq.(\ref{den_ME}). We first note that the  matrix element for unperturbed states $\langle \Psi_{\bold{q},m}^{(0)} |\bar{\rho}_{\bold{q}}| \Psi_{0}^{(0)} \rangle$ scales as $|\bold{q}|^2$ at small $|\bold{q}|$. 
This property follows from the long-wavelength properties of the static structure factor of the unperturbed ground 
state $| \Psi_{0}^{(0)} \rangle$ established in Ref.[\onlinecite{GMP}]:
\begin{equation}
\begin{aligned}
\bar{s}_0(\bold{q})&=\frac{1}{N} \langle \Psi_{0}^{(0)} | \bar{\rho}_{\bold{q}}^\dagger \bar{\rho}_{\bold{q}}  |  \Psi_{0}^{(0)} \rangle \\
&=\frac{1}{N} \sum_{m} |\langle \Psi_{\bold{q},m}^{(0)} |  \bar{\rho}_{\bold{q}}  |  \Psi_{0}^{(0)} \rangle|^2 \\
&\propto |\bold{q}|^4, |\bold{q}| \rightarrow 0.
\end{aligned}
\label{ssf}
\end{equation}
It follows that to first order in both $|\bold{q}|$ and $V_{\bold{G}}$,
\begin{equation}
\begin{aligned}
&\langle \Psi_{\bold{k},m} |\bar{\rho}_{\bold{q}}| \Psi_{0} \rangle 
\approx 
\langle \Psi_{\bold{k},m}^{(0)} |\bar{\rho}_{\bold{q}}| \Psi_{0}^{(1)} \rangle
\\
\approx &\sum_{\bold{G}, n} \delta_{\bold{k},\bold{q}+\bold{G}} 
\frac{V_{\bold{G}} \langle \Psi_{\bold{G},n}^{(0)} |\bar{\rho}_{\bold{G}} | \Psi_{0}^{(0)} \rangle}{E_{0}^{(0)}-E_{\bold{G},n}^{(0)}}
\langle \Psi_{\bold{q}+\bold{G},m}^{(0)} |\bar{\rho}_{\bold{q}}| \Psi_{\bold{G},n}^{(0)} \rangle.
\end{aligned}
\end{equation}
This in turn leads to the following expression of dynamic structure factor:
\begin{equation}
\begin{aligned}
&S(\bold{q},\epsilon)\\
\approx & \frac{1}{N} \sum_{\bold{G},m} \Big[ \Big| \sum_{n} \frac{V_{\bold{G}} \langle \Psi_{\bold{q}+\bold{G},m}^{(0)} |\bar{\rho}_{\bold{q}}| \Psi_{\bold{G},n}^{(0)} \rangle \langle \Psi_{\bold{G},n}^{(0)} |\bar{\rho}_{\bold{G}} | \Psi_{0}^{(0)} \rangle}{E_{0}^{(0)}-E_{\bold{G},n}^{(0)}}
 \Big|^2\\
&\times \delta(\epsilon- (E_{\bold{q}+\bold{G},m}^{(0)} - E_0^{(0)}) )\Big],
\end{aligned}
\end{equation}
where we have neglected the second order correction to the excitation energy  from the potential $\hat{V}$
in the argument of the $\delta$-function.

\begin{figure}[t]
	\includegraphics[width=0.9\columnwidth]{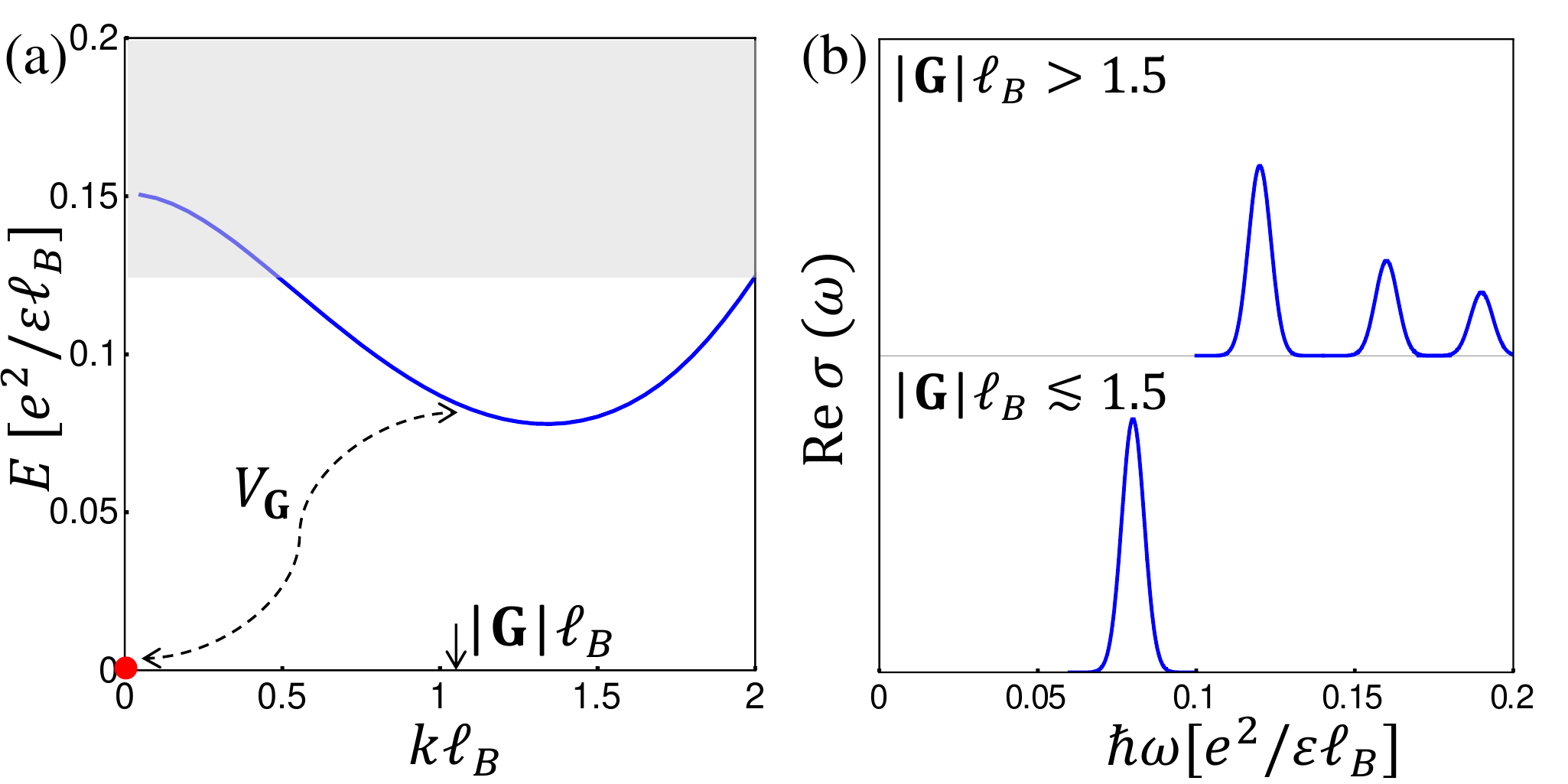}
	\caption{(Color online) Schematic illustration of the perturbation theory analysis. 
	(a)Energy spectrum of Coulomb-only model at filling factor 1/3. The red dot represents the $\nu= 1/3$ Laughlin
	incompressible ground state, the blue line marks the magneto-roton mode and the gray bar the excitation continuum. 
	In the presence of a moir\'e superlattice potential, the perturbed ground state contains  
	admixtures of unperturbed excited states at momenta $\bold{G}$.  This admixture enables intra-LL optical response. (b) Schematic illustration of the optical conductivity. When perturbation theory applies and $|\bold{G}|\ell_B$ is close to 1.5, the SMA is accurate. Thus the intra-LL optical response is dominated by a single peak. When $|\bold{G}|\ell_B>1.5$, weaker 
	optical responses are expected at multiple frequencies. 
	Tuning $|\bold{G}|$ provides a momentum-resolved spectroscopy of 
	FQH excitations.}
	\label{Fig2}
\end{figure}

\section{Single Mode Approximation}
\label{SMA}
To illustrate how moir\'e assisted optical absorption in the quantum Hall regime can 
be interpreted, we focus on the $\nu=1/m$ case for which the fractional quantum Hall gaps are 
largest and the simplifying single mode approximation (SMA) is accurate.
When the SMA applies, a single collective mode exhausts a large fraction of the oscillator 
strength available at a particular wavevector, {\em i.e.} only one state at wavevector $\bold{k}$ 
has a significant value of $\langle \Psi_{\bold{k},n}^{(0)} |\bar{\rho}_{\bold{k}} | \Psi_{0}^{(0)}\rangle$
and that state can therefore be approximated by
\begin{equation}
|\phi_{\bold{k}}^{(0)} \rangle = \frac{1}{\sqrt{N \bar{s}_0(\bold{k})}}\bar{\rho}_{\bold{k}} | \Psi_{0}^{(0)} \rangle,
\end{equation}
where $\bar{s}_0(\bold{k})$ is the static structure factor defined in Eq.(\ref{ssf}).
It follows that the long wavelength dynamic structure factor satisfies 
\begin{equation}
\begin{aligned}
S(\bold{q},\epsilon)
\approx  \sum_{\bold{G}}  \Big|\frac{ V_{\bold{G}} \langle \Psi_{0}^{(0)} | \bar{\rho}_{\bold{q}+\bold{G}}^\dagger \bar{\rho}_{\bold{q}} \bar{\rho}_{\bold{G}}   | \Psi_{0}^{(0)} \rangle }{N \Delta_{\bold{G}} \sqrt{\bar{s}_0(\bold{q}+\bold{G})}}\Big|^2
\delta(\epsilon- \Delta_{\bold{q}+\bold{G}} ),
\end{aligned}
\end{equation}
where $\Delta_{\bold{G}}$ is the Coulomb energy difference between $ |\phi_{\bold{G}}^{(0)}\rangle$ and $|\Psi_{0}^{(0)}\rangle$, 
and 
\begin{equation}
\frac{\langle \Psi_{0}^{(0)} | \bar{\rho}_{\bold{q}+\bold{G}}^\dagger \bar{\rho}_{\bold{q}} \bar{\rho}_{\bold{G}}   | \Psi_{0}^{(0)} \rangle}{N\bar{s}_0(\bold{q}+\bold{G})}
\approx i \ell_B^2 (\bold{q}\times\bold{G})\cdot \hat{z}.
\end{equation}
Therefor the dynamic structure factor to second order in both $V_{\bold{G}}$ and $|\bold{q}|$ is:
\begin{equation}
S(\bold{q},\epsilon)
\approx \frac{1}{2}\sum_{\bold{G}}  \ell_B^4 |\bold{q}|^2 |\bold{G}|^2  \frac{|V_{\bold{G}}|^2 }{\Delta_{\bold{G}}^2}\bar{s}_0(\bold{G})
\delta(\epsilon- \Delta_{\bold{G}}  ).
\label{Spert}
\end{equation}
Applying Eq.(\ref{Resigma}) then yields the following remarkably simple expression
for the real part of optical conductivity: 
\begin{equation}
\text{Re}\sigma(\omega) \, \approx  \, \frac{N}{4N_\phi}\frac{e^2}{\hbar}
\sum_{\bold{G}}  \ell_B^2 |\bold{G}|^2 \frac{|V_{\bold{G}}|^2 }{\Delta_{\bold{G}}}\bar{s}_0(\bold{G})
\delta(\hbar\omega- \Delta_{\bold{G}}  ).
\label{Sigmapert}
\end{equation}
Since linear response of $\langle \bar{\rho}_{\bold{G}} \rangle_0$ to
the moir\'e potential in the SMA is
\begin{equation}
\langle \bar{\rho}_{\bold{G}} \rangle_0 \approx -2 N \frac{ V_{\bold{G}}^* }{ \Delta_{\bold{G}} } \bar{s}_0(\bold{G}),
\end{equation}
Eq.~(\ref{Sigmapert}) satisfies the $f-$sum rule of Eq.(\ref{sum-rule}). 
The perturbation theory and SMA are schematically demonstrated in Fig.~\ref{Fig2}.

\section{Discussion of Experimental Implications}
\label{discussion}
FQH states at filling factors $1/3$, $2/3$, $4/3$ and $5/3$ have been observed in moir\'e-patterned graphene on a hBN substrate using 
capacitance\cite{Hunt13} and transport\cite{Dean15} measurement.  Our theory predicts that if light absorption 
measurements were performed in these samples, they would have a finite intra LL signal,
providing the first truly spectroscopic probe of fractional quantum Hall collective excitations.
Intra-LL collective excitations have a typical energy $\sim 0.1 e^2/ (\varepsilon \ell_B)$, 
which is about $10$ meV  ( in the THz frequency range ) at 35T if 
we use $\varepsilon = 3.5 $ for the effective dielectric constant. 
In the SMA  [Eqs.~(\ref{Spert}) and (\ref{Sigmapert})] the excited states at wave 
vectors $\bold{G}$ saturate the $f-$sum rule.
The SMA is particularly accurate for the Laughlin state when $\bold{G}$ is close to the wave vector of the 
magneto-roton minimum, and perturbation theory requires
$|V_{\bold{G}}|$ to be small compared to 
$\Delta_{\bold{G}}$.  For $\nu = p/3$ Laughlin states, the roton minimum 
is located around $k \ell_B \sim 1.5$.
In aligned graphene/hBN the moir\'e pattern has a period of 14nm. 
$|\bold{G}|\ell_B$ is then about 2.2 at 35T, exceeding the value at which the single-mode-approximation 
is most accurate and opening a door to the poorly understood crossover between 
magneto-roton collective modes and fractional particle-hole excitations.  

The FQH effect in graphene is enriched by spin and valley degrees of freedom\cite{Wu14, Wu15}.  
%Spin and valley are expected to be polarized for the FQH states at $\nu = p/3$ 
%because of Zeeman energy and 
%sublattice dependent potential, which is further reinforced by electron interaction effect. 
For $N=0$ LLs, electron states in opposite valleys are localized on opposite sublattices.  
Because the moir\'e potential of graphene on hBN\cite{Jung15} is sublattice dependent
when projected onto the lowest Landau level, THz absorption could also 
be used to detect valley polarization.  
%The $f-$sum rule in Eq.~(\ref{sum-rule}) can be readily applied to FQH states in graphene/hBN. 

Twisted TMD bilayers are another candidate for moir\'e assisted FQH spectroscopy.  
Common chalcogen TMD heterjunctions, for example WSe$_2$/MoSe$_2$, 
have particularly long period moir\'e superlattices when aligned.
The lattice constants of WSe$_2$ and MoSe$_2$ have a mismatch of only $0.1\% \sim 0.2\%$\cite{Xiao12}, much smaller that for 
graphene/hBN.   The recent observation of Shubnikov-de Haas oscillations and 
quantum Hall states\cite{Tutuc16} in high mobility holes in monolayer WSe$_2$ promises the future 
realization of FQH states.  Some differences between 
TMD bilayers and graphene/hBN could prove interesting because:
(1)The N=0 hole LLs of WSe$_2$ and MoSe$_2$ have neither spin nor valley\cite{Li13} degeneracy, 
due to a combination of broken inversion symmetry and strong spin-orbit coupling,\cite{Xiao12}
simplifying theoretical models and the interpretation of any signals that are observed. 
(2)The moir\'e potential is expected to be weaker\cite{Wu16}, providing stronger justification for the perturbative 
interpretation we propose, because a TMD monolayer consists of three atomic layers with low-energy electrons
located primarily in the middle layer. 
The moir\'e potential in TMDs can be tuned by an electric displacement field that is perpendicular to the bilayers.
(3)  Because of the small lattice constant mismatches of common chalcogen TMD heterjunctions
$|\bold{G}|\ell_B$ can be tuned across the roton minimum of
the $1/3$ Laughlin using convenient twist angles.  For example using a twist angle of 
$\sim 0.9^{\circ}$ between WSe$_2$/MoSe$_2$ in a 35T magnetic field, $|\bold{G}|\ell_B$ is close to 1.5.  This property
should allow the roton minimum dispersion to be measured accurately. 

We have assumed that the moir\'e potential is weak so that FQH states survive. When such assumption breaks down, there can be phase transitions between FQH and Wigner crystal states.\cite{Ashley}
 We have mainly focused on Laughlin states. It will be at least equally
 interesting\cite{Son2015, Geraedts2106} to probe FQH states at filling factors close to 1/2 optically,
 since these states are less well understood theoretically.
\section{Acknowledgment}
Work at Austin was supported by the Department of Energy, Office of Basic Energy Sciences under contract DE-FG02-ER45118 and by the Welch foundation under
grant TBF1473.  The work of FW at Argonne National Laboratory
was supported by the Department of Energy,
Office of Science, Materials Sciences and Engineering Division.


\begin{thebibliography}{}
\bibitem{sublattice}
%Because graphene is a gapless semiconductor, the 
%low-energy Hamiltonian of graphene must describe two-bands, corresponding to the two
%honeycomb sublattice sites.  
The moir\'e Hamiltonian for graphene on hBN is a spatially periodic $ 2 \times 2$
matrix operator that acts on sublattice degrees of freedom.  See for example Ref.~\onlinecite{Jung14}.  However, because lowest LL wavefunctions exist purely on one sublattice only the potential on that sublattice is relevant for FQH physics.
\end{thebibliography}
\end{document}